\documentclass{aa}

\usepackage{graphicx}
\usepackage{txfonts}
\usepackage[colorlinks=true,     linkcolor=blue, citecolor=blue, filecolor=blue, urlcolor=blue]{hyperref}
\begin{document}

 \title{Unraveling the untwisting process and upward\\
  mass transfer of a twisted prominence driven by vortex motion}

   \author{X. F. Zhang
          \inst{\ref{inst1},\ref{inst2},\ref{inst3}}
          \and
          G. P. Zhou\inst{\ref{inst1},\ref{inst2},\ref{inst3}}
          \and
          C. L. Jin
          \inst{\ref{inst1},\ref{inst2},\ref{inst3}}
          \and
          Y. Z. Zhang
           \inst{\ref{inst1},\ref{inst2},\ref{inst3}}
          \and
          G. W. Li
           \inst{\ref{inst1},\ref{inst3}}
          \and\\
          Z. H. Shang
           \inst{\ref{inst4}}
          \and
          L. P. Li
           \inst{\ref{inst1},\ref{inst2},\ref{inst3}}
          \and
          S. B. Yang
           \inst{\ref{inst1},\ref{inst2},\ref{inst3}}
          \and
          S. H. Yang
           \inst{\ref{inst1},\ref{inst2},\ref{inst3}}
          \and
          J. X. Wang
           \inst{\ref{inst3},\ref{inst1},\ref{inst2}}
          }

   \institute{National Astronomical Observatories, Chinese Academy of Science, A20 Chaoyang District, Beijing, 100101, China \label{inst1}\\
              \email{gpzhou@nao.cas.cn}
         \and
             State Key Laboratory of Solar Activity and Space Weather, Beijing 100190, China \label{inst2}
	\and
	    University of Chinese Academy of Sciences, Beijing 100049, China \label{inst3}
	 \and
	    Yunnan Key Laboratory of Artificial Intelligence, Faculty of Information Engineering and Automation, Kunming University of Science and Technology, Kunming, 650500, China \label{inst4}
             }

   \date{Received 26 September 2023; accepted 7 July 2024}


\abstract
   {Solar filaments or prominences are common features in the Sun's atmosphere that contain cool chromospheric material suspended within the hot corona. However, the intricate topology of these structures and the mechanisms driving their instability and upward material transfer are not well understood. Investigating these issues is essential for gaining insight into the fundamental laws that govern solar activity.}
   {This study is to analyze a specific twisted prominence observed on February 10, 2021, and to explore its dynamics, including stability, motion, and material transfer. The study also aims to propose a mechanism, based on the K\'arm\'an Vortex Street instability, to explain the destabilization of the prominence.}
   {The study utilizes high-resolution H$_{\alpha}$ observations from the 1-m New Vacuum Solar Telescope and space-borne observations from the Solar Dynamics Observatory. These observations capture the characteristics and behavior of the twisted prominence. We analyzed the data to investigate the equilibrium state, subsequent destabilization, vortex motion, oscillations, resonations, untwisting, and upward mass loading of the prominence. We also detected and measured the speeds of outflows surrounding the prominence.}
   {The study reveals that the observed twisted prominence exhibited a stretched and twisted structure at its apex, distinguishing it from familiar cloudy prominences. Following a period of more than 30 hours in equilibrium, the prominence underwent destabilization, leading to a series of dynamic phenomena, such as vortex motion, oscillations, resonations, untwisting, and the upward transfer of mass. Consequently, material from the top of the prominence was carried upward and deposited into the overlying magnetic arcades. Noteworthy, outflows surrounding the prominence were characterized by speeds exceeding 40 km s$^{-1}$.}
   {Based on these findings, we propose, for the first time, a mechanism rooted in the K\'arm\'an Vortex Street instability to explain the destabilization of the prominence. The estimated typical Strouhal Number of 0.23$\pm$0.06, which is related to vortex shedding, falls within the expected range for the K\'arm\'an Vortex Street effect, as predicted by simulations. These discoveries provide new insights into the dynamics and fundamental topology of solar prominences and reveal a previously unknown mechanism for mass loading into the upper atmosphere.}


 \titlerunning{Unraveling Dynamics of a Twisted Prominence Driven by Vortex Motion}
   \authorrunning{Zhang et al.}
   \maketitle

%

\section{Introduction}
Filaments, also known as prominences, are characterized as dense and cold plasma, whose suspension in the hot corona is facilitated by magnetic fields. Usually, dark filaments are typically observed in absorption against the solar disk, while bright prominences are detected in emission at the limb. These structures, fundamental to our understanding of the Sun, have been extensively studied throughout history, both through observations and theoretical approaches. They serve as a key subject in the field of solar physics \citep[e.g.][and references therein]{2010SSRv..151..333M}, as their physics holds the key to unraveling crucial components of solar activity and magnetic structure on the Sun \citep[e.g.][]{1998ASPC..150.....W}. The advent of advanced observations provides an excellent opportunity to gain a deeper understanding of the intricate physical processes associated with prominence activity, prompting significant research efforts in this area.

Many reviews contribute to our current understanding of the structures and dynamics of filaments and prominences\citep[e.g.][]{1998SoPh..182..107M, 2010SSRv..151..333M}. These features are consistently observed along filament-channels \citep[e.g.][]{2010SSRv..151..333M}, typically aligned with polarity inversion lines \citep[PILs,][]{1955ApJ...121..349B} that separate regions of positive and negative magnetic polarities on the photosphere. In the vicinity of PILs, various photospheric motions frequently occur, such as magnetic flux emergence \citep[][]{1993SoPh..143..119W, 2000SoPh..195..347D, 2007SoPh..244...13Z}, flux cancellation\citep[e.g.][]{1989SoPh..121..197L, 2016SoPh..291.2373Z}, and wave disturbances \citep[e.g.][]{2014LRSP...11....1P}, among others. These motions significantly shear the magnetic fields, giving rise to intricate and complex prominence structures that are predominantly devoid of vertical fields \citep[e.g.][]{2003SoPh..216..143W}.

Filaments or prominences suspend in the corona exhibit highly dynamic activity, including plasma downflows \citep[][]{1981SoPh...70..315E}, vortices of approximately 10$^{5}$ km $\times$ 10$^{5}$ km in size \citep[][]{1984SoPh...91..259L}, counterstreaming flows \citep[][]{1998Natur.396..440Z}, plumes, plasma bubbles or blobs, finger-like or mushroom-like structures and more \citep[e.g.][]{2010A&A...514A..68S, 2011Natur.472..197B}. While filaments or prominences can maintain global stability in the corona for weeks \citep[][]{1986NASCP2442..149H}, they can also undergo sudden eruptions, losing equilibrium and resulting in dramatic events  \citep[e.g.][]{2006ApJ...651.1238Z}. Thus, understanding the various instability mechanisms driving these phenomena is not only essential for deciphering the fundamental mechanism of solar activity but also for predicting solar storms.

With the advancement of high-resolution observations, modern techniques allow us to uncover intricate internal structures like bubbles or plumes within the body of a prominence, especially in the bridge-pier-like bottom. These structures display a range of unstable mechanisms. For instance, Rayleigh-Taylor instability \citep[RTI,][]{2005Natur.434..478I} is frequently observed in the plasma bubbles of quiescent prominences \citep[e.g.][]{2021A&A...646A..93P}. Additionally, the magnetic RTI is proposed as the driving mechanism behind rising plumes that carry hot plasma and magnetic flux upward through the prominence \citep[e.g.][]{2011Natur.472..197B}.

 Dynamic motions occurring within solar prominences can also induce the Kelvin-Helmholtz Instability \citep[KHI,][]{2018ApJ...864L..10H}. Using H$_{\alpha}$ red-wing observations at +0.3 {\AA} from the New Vacuum Solar Telescope \citep[NVST][]{2014RAA....14..705L, 2020ScChE..63.1656Y}, \citet[][]{2018ApJ...863..192L} identified KHI in surface or interface of prominence plumes as rapidly evolving vortex-like structures with counterclockwise or clockwise rotations. It also suggests that KHI can contribute to plume formation through the strong shear flow at the bubble-prominence interface \citep[e.g.][]{2017ApJ...850...60B}. The shear flow velocities can reach 10 km s$^{-1}$ above the bubble, and up to 100 km s$^{-1}$ within the interior. In \citet[][]{2018ApJ...857..115Y}, KHI is associated with the oscillation of the stream guided by the magnetic field and the surrounding medium. Furthermore, some authors discuss that KHI, in its nonlinear phase, can transport prominence material into the upper layers and trigger plasma ejections \citep[e.g.][]{2011ApJ...729L...8F}. Under specific conditions, unstable plumes caused by a mixing hybrid of KH-RT instabilities may lead to the eruption of a loop-like prominence \citep[][]{2019ApJ...874...57M}.

 In addition to the mentioned instabilities, various other instabilities are studied and modeled in the context of prominence dynamics. The ideal kink instability is one of the earliest discovered instabilities that destabilize the flux system within prominence \citep[e.g.][]{1996ApJ...464L.199R, 2014ApJ...789L..35C}. Using simple analytical methods to model the plasma state within solar prominences, dissipative instabilities are identified in different interfaces, such as those between incompressible and partially ionized prominence plasma \citep[][]{2015A&A...577A..82B}, or between two partially ionized compressible plasmas \citep[][]{2012ApJ...749..163S}. Thermal instability and thermal nonequilibrium are frequently discussed in the dynamics of prominences \citep[e.g.][]{2011Natur.472..197B, 2019SoPh..294..173K}. The process of catastrophic radiative cooling can lead to the formation of dense plasma concentrations, which break the frozen-in flux condition and give rise to discrete downflow blobs or droplets \citep[][]{2012ApJ...755...34L, 2016ApJ...823...22X}. These phenomena highlight the importance of thermal processes in the overall behavior of prominences.

 In this work, our focus is on unveiling a new instability known as the K\'arm\'an vortex street\citep[][]{wille60}, which led to the disintegration of a fascinating twisted prominence. The K\'arm\'an vortex street is a phenomenon where two swirling vortices are formed due to the unsteady separation of fluid flow around blunt bodies, and they oscillate together, causing the alternated separation of the blunt bodies \citep[][]{2000ifd..book.....B}. Evidence of vortex shedding and its role in the oscillation of post-flare loops has been proposed by \citet[][]{2019PhRvL.123c5102S}. \citet[][]{2009A&A...502..661N} modeled vortex shedding to explain the excitation of kink-mode oscillations in coronal loops via upflows. In our study, we discover the novel effect of the K\'arm\'an vortex street in destabilizing a twisted cloudy prominence. The observational results and findings are presented in \S 2, while the discussions and conclusions can be found in \S 3.


\section{Observations and results} \label{sec:obs}

 A bridge-pier-like prominence initially appeared at the southeast edge of the Sun around 10:00 UT on 2021 February 8. By February 9 at 18:00 UT, the twisted cloudy prominence began to display signs of instability. The dynamic evolution of this prominence between 03:35-08:36 UT on February 10 was observed using the H$_{\alpha}$ imager on the NVST with a high spatial resolution of  0.165$''$. The entire physical process of the prominence was also captured by the Extreme Ultraviolet (EUV) passbands of the Atmospheric Imaging Assembly \citep[AIA;][]{2012SoPh..275...17L} on board the Solar Dynamics Observatory \citep[SDO;][]{2012SoPh..275....3P}. The AIA observations has a cadence of 12 s and a spatial resolution of approximately 0.6$''$. This prominence was part of a larger filament situated along a PIL, which was revealed by the magnetograms obtained from the Helioseismic and Magnetic Imager \citep[HMI,][]{2014SoPh..289.3483H} on the SDO. The HMI magnetograms provide a time cadence of 45 s and a spatial resolution of 0.5$''$. Additionally, H$_{\alpha}$ observations from the Solar Magnetism and the Activity Telescope at Huairou Solar Observing Station \citep[HSOS;][]{2007ChJAA...7..281Z} detected the long filament and the prominence within its global field of view. All the H$_{\alpha}$ data were coordinated with the SDO field of view at 03:35 UT on 2021 February 10.

\begin{figure}

\resizebox{\hsize}{!}{\includegraphics{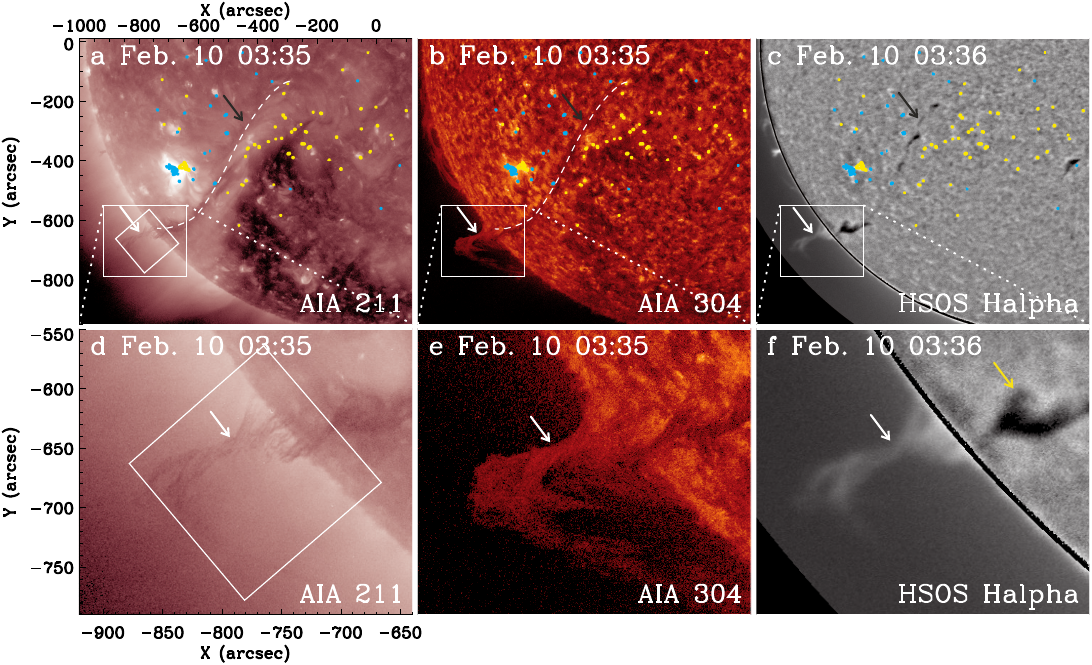}}
\caption{Overview of the twisted prominence observed on 2021 February 10.  The prominence is situated at the southeast edge of the Sun, as indicated by white arrows in the EUV images at 211 {\AA} (Panels a and d) and 304 {\AA} (Panels b and e) around 03:35 UT. It is a component of a larger filament, as shown by a black arrow in Panel (c), which is a H$_{\alpha}$ image captured by HSOS at 03:36 UT. The dashed curves in Panels (a-b) depict the PIL that separates positive and negative flux patches. To highlight the twisted structure of the prominence, sub-images in the larger squares of Panels (a-c) are displayed in Panels (d-f). The cyan and yellow outlines represent positive and negative flux density at $\pm$100 G, respectively. The long filament is indicated by dashed lines in Panels (a-b). The slanted squares in Panels (a, d) indicate the field of view of  H$_{\alpha}$ observations from NVST. \label{fig:f1}}
\end{figure}

\subsection{Twisted structure and dynamic evolution of the prominence based on multi-passband observations} \label{sub31}

The examined prominence represents one of the bridge-pier-like structures found within a long filament. The comprehensive AIA and HMI observations from SDO provide a clear view of the corresponding long filament situated within a PIL. When viewed in a smaller field of view, the prominence exhibits a twisted structure and displays a cloudy appearance in most of the EUV passbands (see Figure \ref{fig:f1}d), except for the EUV 304 {\AA}, which shows an optically thick characteristic (Figure \ref{fig:f1}b, \ref{fig:f1}e). In the full-disc H$_{\alpha}$ observations captured by HSOS, the long filament appears as a dark feature when absorbed by the solar disk, while the prominence is clearly visible as a bright emission at the east limb (Figure \ref{fig:f1}c). The long dark filament is supported by multiple bridge-pier-like prominences, some of which may display a dark feature against the disk (indicated by the yellow arrow in Figure \ref{fig:f1}f). The complete evolution of the prominence has been monitored through EUV observations from AIA since 2021 February 8.

\begin{figure}
\resizebox{\hsize}{!}{\includegraphics{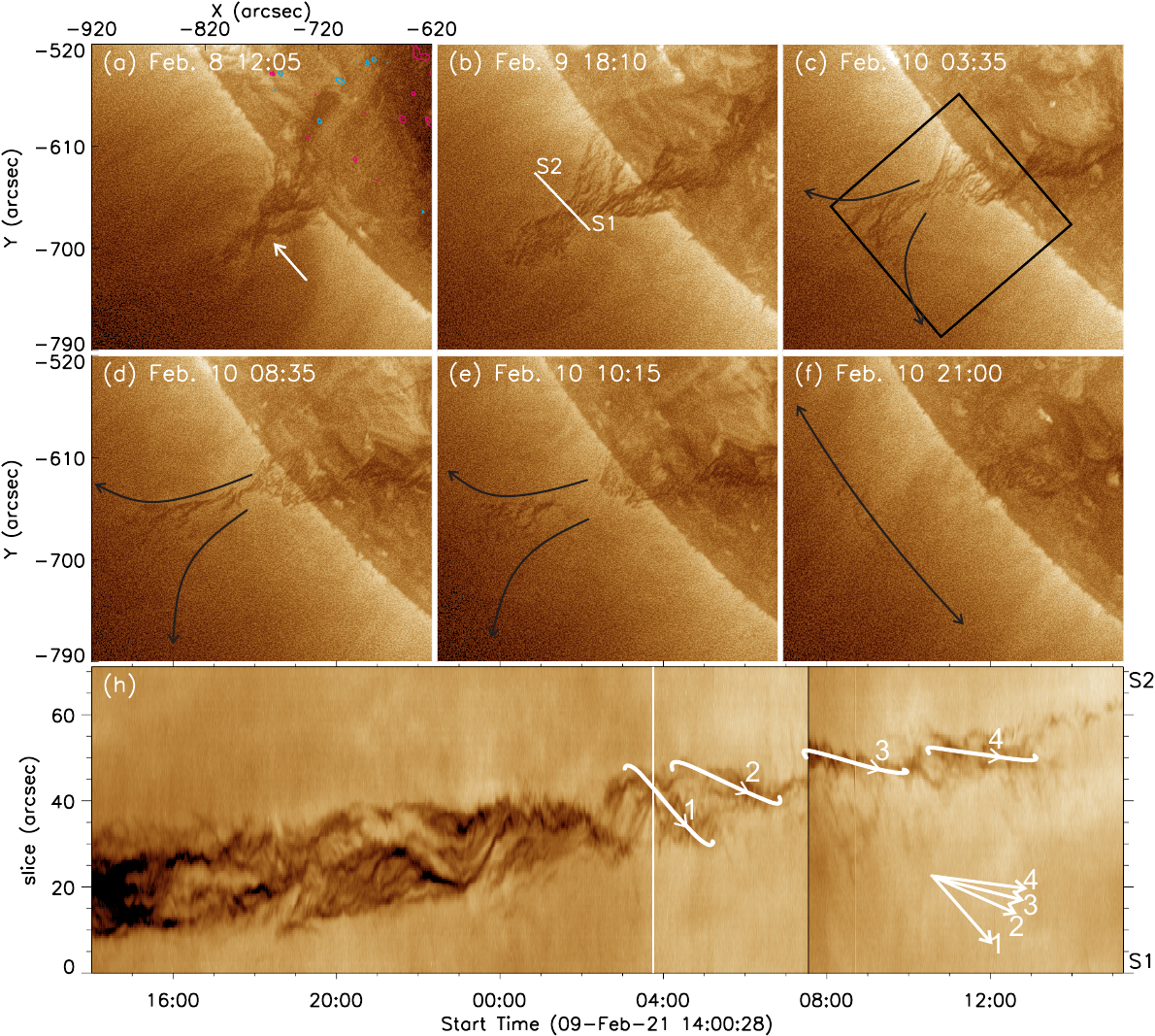}}
\caption{Evolutions of the prominence captured by EUV observations at 193 {\AA} from 2021 February 8 to 10. Initially, on February 8, the prominence is observed at the southeast edge, exhibiting a twisted structure with a bridge-pier body (Panels a-e). The twisting neck of the prominence is indicated by the white arrows. From 18:10 UT on February 9, material starts to move upwards from the top of the prominence towards the north and the south, as shown by the black arrows in Panels (c-e). At 21:00 UT on February 10, the top twisted structure of the prominence begins to diminish, while a new dark arcade appears above the prominence, highlighted by the black double-head arrow in Panel (f). EUV intensity distributions along the slice "L1-L2", as indicated in Panel (b), are depicted in Panel (h). The white winding lines labeled "1-4" highlight the twisting structures of the prominence around its axis. In the lower right corner, arrows "1-4" illustrate the decreasing angles between the twisting features and the prominence axis. The long PIL is depicted by the red dashed lines. The black box in Panel (c) represents the field of view of the H$_{\alpha}$ observations from NVST, as illustrated in Figure \ref{fig:f3}. \\ (An animation of corresponding EUV observations is available.)\label{fig:f2}}
\end{figure}

The dynamic evolution of the twisted prominence is extensively documented through uninterrupted EUV observations using AIA from 2021 February 8 to 10. The prominence initially displays a twisted structure and appeared at the southeast edge on February 8 (Figure \ref{fig:f2}a). Maintaining equilibrium for approximately 30 hours, it then becomes unstable at around 18:10 UT on February 9 (Figure \ref{fig:f2}b). Subsequently, 9 hours later at approximately 03:35 UT on February 10, the prominence exhibits untwisting characteristics, accompanied by the stretching and thinning of its twisted neck (indicated by the white arrows in Figure \ref{fig:f2}a). Over time, dark material from the top of the prominence gradually moves upwards, spreading towards the north and south sides (as illustrated by the two black arrows in Figure \ref{fig:f2}c-\ref{fig:f2}e). Ultimately, the twisted top of the prominence vanishes, leaving behind a transverse dark arcade (Figure \ref{fig:f2}f). Figure \ref{fig:f2}g illustrates the EUV intensity distributions along the slice "L1-L2", as indicated in Figure \ref{fig:f2}b. The dark winding features, highlighted by white curves "1-4", represent the twisting threads of the prominence. Following ~3:35 UT on February 10, the winding threads exhibit a tendency to incline towards the prominence axis. This observation aligns with the expected untwisting process of the prominence. Detailed observations conducted by NVST using fine H$_{\alpha}$ imaging will shed light on the intricate mechanisms underlying the disturbance and untwisting of the prominence.

\subsection{Vortex motion responsible for untwisting the prominence and causing upward transfer of mass\label{sub33}}

\begin{figure}
\resizebox{\hsize}{!}{\includegraphics{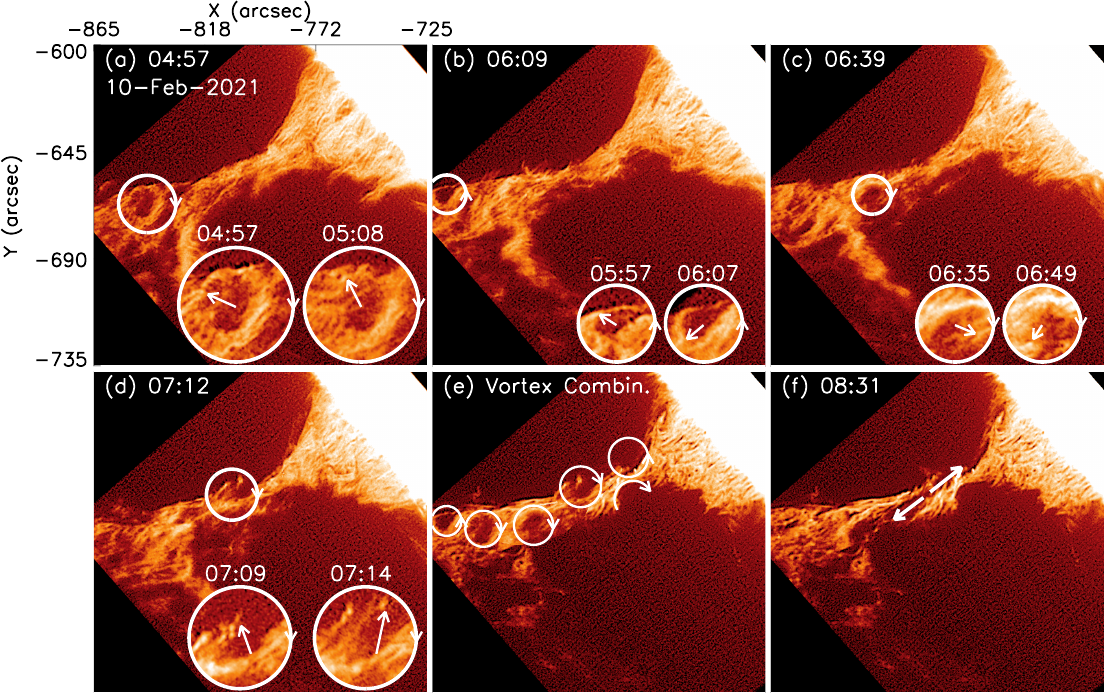}}
\caption{H$_{\alpha}$ observations at the line center of 6563 {\AA} from NVST showing the vortex motion on both sides of the prominence and its resulting effects. Along the prominence, as depicted in Panels (a-f), we identify several vortices at different positions, which are highlighted using white circles to demonstrate their diverse rotating directions. To determine the rotating directions, each selected vortex is magnified two times at the bottom of Panels (a-d), and the movements of brightening features are tracked as shown by white arrows. Furthermore, in Panel (e), the vortices from Panels (a-d) are superimposed on the image at 08:31 to illustrate a sequential vortex process. Finally, Panel (f) displays the redistribution of prominence material, indicated by two arrows pointing in opposite directions.\\
(An animation of this figure is available.)\label{fig:f3}}
\end{figure}

H$_{\alpha}$ observations from NVST capture the dynamic and unstable evolution of the prominence between 03:45 UT and 08:34 UT on February 10. In the H$_{\alpha}$ observations as shown in Figure \ref{fig:f3}, vortex movements can be identified at different moments on both sides of the prominence in the plane of the sky, as denoted by white circles with arrows indicating rotation directions. Due to projection effects within the complex twisting structure of the prominence, it is not possible to directly observe an ideal successive vortex motion. However, when the vorticities identified at different moments are synthesized, a pattern of successive vortex movements can be observed, as depicted in Figure \ref{fig:f3}e. The five vortices highlighted by circles in Figure \ref{fig:f3}e, from left to right, have diameters of 8.4 Mm, 6.5 Mm, 10.4 Mm, 11.0 Mm and 6.7 Mm, respectively. These vortices exhibit a sense of continuity and are significantly larger compared to those discussed in the bridge-pier-like base of a prominence, which typically have an average size of about 3 Mm \citep[e.g.][]{2008ApJ...676L..89B, 2018ApJ...857..115Y}. The larger vortices are believed to be a result of the other instability different from RTI or KHI as discussed in \citet[][]{2008ApJ...676L..89B} or \citet[][]{2018ApJ...857..115Y}. During this evolution, the prominence underwent stretching, untwisting, and thinning of its twisted neck region. At approximately 08:31 UT, a significant redistribution of mass occurs at the top of the prominence. Material from the top of the prominence exhibited continuous upward transfer, with flows directed towards both the north and south directions. Eventually, the raised portion of the prominence became untwisted from the lower bridge pier, as indicated by the two white arrows in Figure \ref{fig:f3}f.

\begin{figure}
\resizebox{\hsize}{!}{\includegraphics{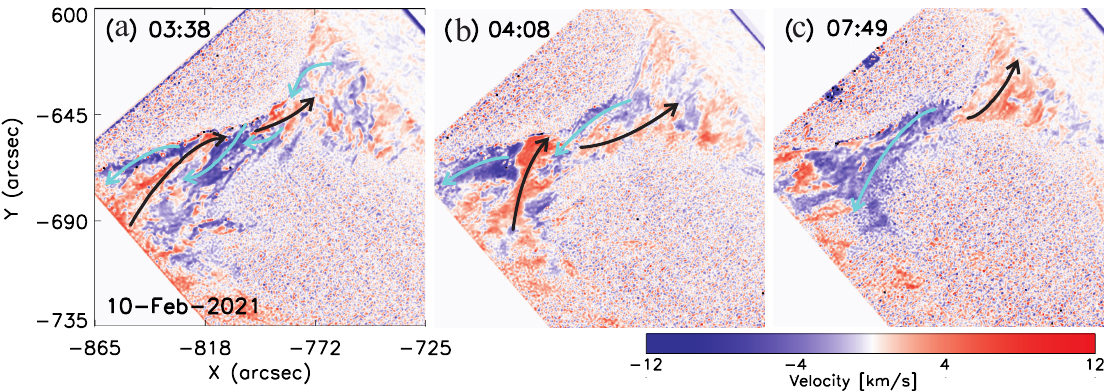}}
\caption{Doppler response revealing the untwisting of the prominence. In Figure \ref{fig:f4}, cyan arrows highlight blue shifts, indicating mass motions toward us, while black arrows mark red shifts, representing movements away from us. The bundles of blue and red shifts exhibit a twisting pattern, completing at least 2.5 cycles at 03:38 UT (Panel a), and gradually decreasing to as low as 1 cycle by 07:49 (Panel c).\label{fig:f4}}
\end{figure}

The dynamic twisting of the prominence can be revealed through the Doppler velocity distributions obtained from the H$_{\alpha}$ observations, as illustrated in Figure \ref{fig:f4}. Taking into account the weighting of the H$_{\alpha}$ line center and line offsets, the Doppler velocities are calculated using the formula: $\frac{3\times10^{5}\times0.3\times(I_{r}-I_{b})}{(I_r+I_b+I_0)\times6563}$, where $\emph{I}_{0}$ represents the line center at 6563 {\AA}, $\emph{I}_{r}$ denotes the line offset at 0.3 {\AA}, indicating a red shift, and $\emph{I}_{b}$ represents the line offset at -0.3 {\AA}, indicating a blue shift. The blue and red Doppler shifts observed correspond to plasma motion within the prominence threads. Figure \ref{fig:f4} illustrates this phenomenon with cyan and black arrows, representing the twist-like distributions of blue and red shifts, respectively. These distributions reflect the twist features present in the prominence structure. Initially, the twists are observed to have at least 2.5 cycles at 03:38 UT in Figure \ref{fig:f4}a, reaching approximately one cycle by 07:49 UT in Figure \ref{fig:f4}c. This gradual reduction in twist cycles indicates an untwisting process occurring within the prominence.

\begin{figure}
\resizebox{\hsize}{!}{\includegraphics{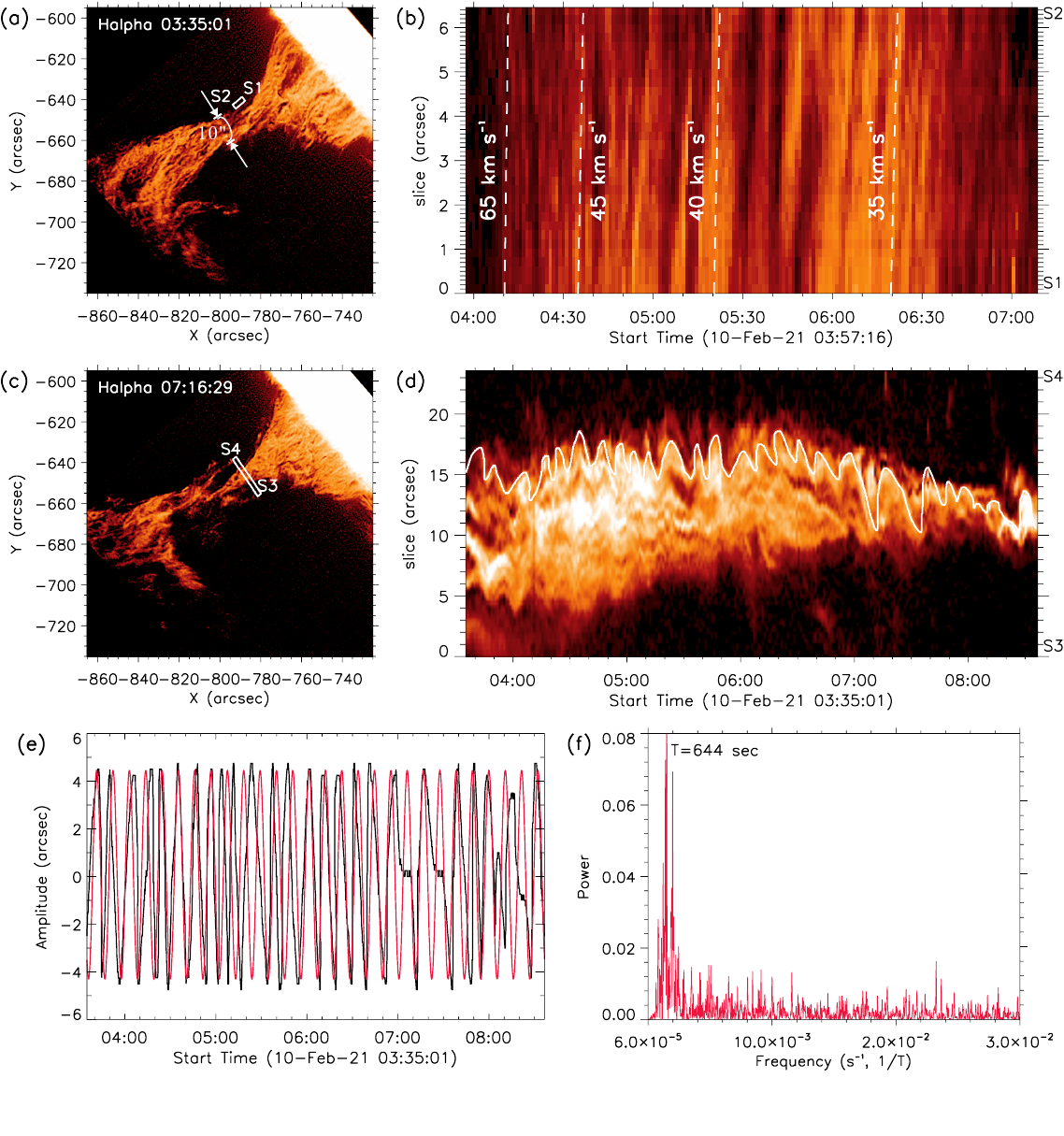}}
\caption{Outflow speeds and dominant oscillation of the prominence calculated based on the time-slice H$_{\alpha}$ intensity distributions. Panels (a) and (c) show the rectangle slices "S1S2" and "S3S4" in the H$_{\alpha}$ images at 03:35 UT and 07:16 UT, respectively. Panels (b) and (d) present the corresponding time-slice intensity distributions from "S1" to "S2" and from "S3" to "S4", with time represented on the X-axis and distances on the Y-axis. The dashed lines in Panel (b) indicate the linear fitting of the outflows. In Panel (d), the white curve highlights a notable oscillation along the "S3-S4" line. This oscillation is depicted by the black curve in Panel (e), which is obtained by subtracting the smoothed intensity with a width of 91. The red curve in Panel (e) represents the fitting curve computed using the Lomb-Scargle periodogram. Panel (f) displays the fitting frequency and power, indicating a dominant period of 644 seconds.\label{fig:f5}}
\end{figure}

H$_{\alpha}$ observations reveal periodic vortex motion occurring during the unstable process of the prominence. Determining the dominant period caused by these vortex movements is crucial for understanding the instability mechanism of the prominence. In Figure \ref{fig:f5}c, a region within the prominence that experienced the most violent vortex motion was selected by placing a rectangle slice across it. Initially, multiple oscillations are observed in the time-distance distributions of H$_{\alpha}$ intensity, as depicted in Figure \ref{fig:f5}d. As time progresses, these distinct oscillations gradually merge into a new pulsation with a larger amplitude around 07:00 UT. This merging suggests that the prominence has entered a state of resonance, which could explain its untwisting, stretching, and thinning of the twisting neck.

To determine the typical period of the prominence oscillation, the Lomb-Scargle periodogram method was applied using an automatically determined frequency grid \citep{2018ApJS..236...16V}. The white curve in Figure \ref{fig:f5}d represents the highlighted oscillation period. Based on the fitting result shown in Figure \ref{fig:f5}e-\ref{fig:f5}f, the typical oscillation period during the unstable process of the prominence is determined to be 644 seconds.

\section{Discussions and conclusions} \label{sec:dis}

\begin{figure}
\resizebox{\hsize}{!}{\includegraphics{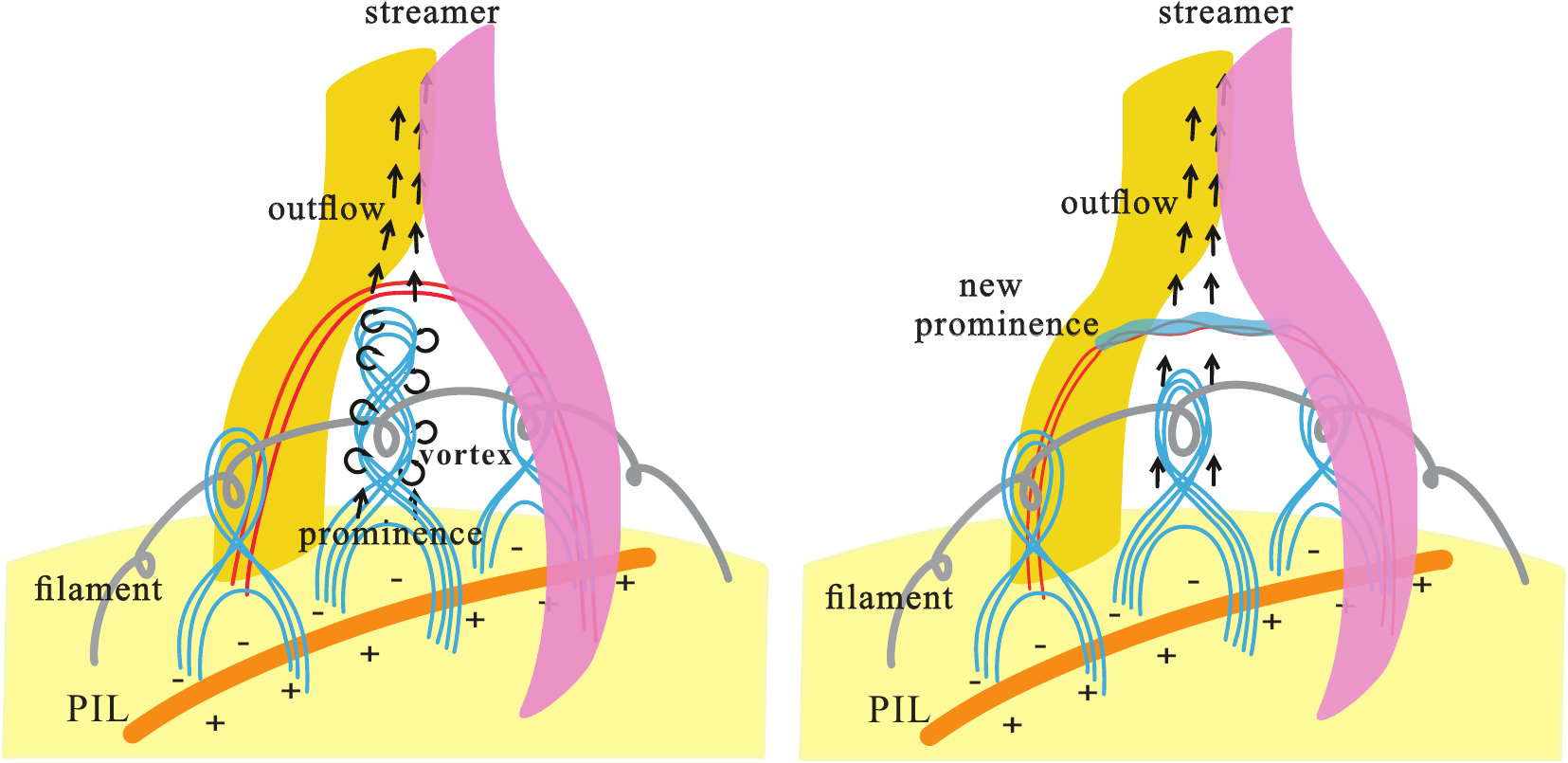}}
\caption{Cartoons illustrating the vortex motion that occurs when outflows in the streamer interact with the twisted prominence. The magnetic structures depicted from the bottom to the top include the PIL indicated by the yellow line, multiple bridge-pier-like prominences represented by the blue curves, overlying arcades shown in red, and the streamer shaded in pink and yellow. The straight black arrows represent the plasma propagating upward along the open magnetic fields in the streamer. The curved arrows indicate the outflow being obstructed by the prominence, leading to the formation of vortex motion. As the prominence untwists, its mass at the apex is redistributed up toward the red magnetic arcades, as illustrated in the right panel. \label{fig:f6}}
\end{figure}

The analysis reveals that the prominence undergoes an instability that is characterized by a repeating pattern of swirling vortices, eventually leading to the untwisting of the prominence and mass transfer. These phenomena support the concepts of the K\'arm\'an vortex street effect caused by vortex shedding, a process responsible for the unsteady destruction of a blunt body \citep[][]{von31}. The prominence is situated at the base of a streamer, oriented orthogonal to the PIL, as illustrated in the cartoon of Figure \ref{fig:f6}. Along the open magnetic field lines of the streamer, plasma can propagate outwards, similar to solar winds moving away from the solar disk. The bridge-pier-like prominence, positioned across the PIL, acts as a blunt body obstructing the outward flow, creating favorable conditions for the occurrence of the K\'arm\'an Vortex Street effect. The presence of large and continuous vortices observed away from the prominence in this study strongly indicates the occurrence of K\'arm\'an Vortex Street.

To further explore the unstable mechanism, we estimated the Strouhal number (St), which is suggested as one of the theoretical criteria for the K\'arm\'an vortex street effect \citep[][]{ben26}. St can be calculated using the formula {\it St}={\it L}/{\it P}{\it v}, where  {\it L} represents the size of the blunt body, {\it P} is the period of the vortices, and {\it v} is the velocity of the fluid passing by. According to previous simulations, {\it St} falls within the range of [0.15, 0.25] for the occurrence of K\'arm\'an Vorter Street effect \citep[][]{2010PhRvL.105e5004G}.

In Figure \ref{fig:f5}a, the primary width of the untwisting neck, indicated by the white arrows, is considered as the size of the blunt body and measured to be approximately 10 arcsec, i.e. 7.25$\times10^{3}$ km. The outward flow along the open magnetic field lines of the streamer can be estimated by analyzing the time-slice intensity distribution in Figure \ref{fig:f5}b. It is observed to range in [35, 65] km $s^{-1}$, with a mean value of 50 km $s^{-1}$. By combining this information with the measured period of 644 seconds during the instability of the prominence, we calculate the St to be 0.23$\pm$0.06. This falls within the predicted range of [0.15, 0.25] for the occurrence of the K\'arm\'an Vortex Street effect, as proposed by the previous MHD simulations.

The cartoons presented in Figure \ref{fig:f6} illustrate the occurrence of the K\'arm\'an Vortex Street instability and its impact on the disintegrating of the prominence. Above the PIL, there is a streamer structure with open magnetic fields represented by top yellow and pink shades. The twisted prominence is located at the base of the streamer, oriented orthogonal to the PIL, and covered by closed magnetic arcades indicated by red closed lines. Outward plasma along the streamer's open fields is expected to continuously propagate away from the solar disc, as denoted by the straight arrows. However, the prominence obstructs this plasma outflow, leading to the occurrence of the K\'arm\'an Vortex Street instability. This instability is manifested by the development of vortices along the opposing sides of the prominence, which is further confirmed by quantitative calculations, yielding a Strouhal number of 0.23$\pm$0.06, consistent with simulations.

During the K\'arm\'an vortex street process, the alternating vortices create a difference in instantaneous pressure on the opposing sides of the prominence, resulting in the vibration of the prominence. Initially, the various internal structures within the prominence vibrate differently, as illustrated in Figure \ref{fig:f5}d. However, these distinct vibrations eventually merge into a new oscillation with a large amplitude, indicating a resonating process within the prominence. This resonance is proposed as an important factor in untwisting the prominence and driving the transfer of mass from the prominence's top to the overlying magnetic arcades in the upper atmosphere. It is worth noting that the K\'arm\'an vortex street effect is not only a well-known phenomenon in fluid dynamics, often associated with the collapse of structures like bridges, but is also revealed to occur during magnetohydrodynamic processes, as discussed in this study. It can destabilize the prominence and facilitate the upward transfer of material in the solar atmosphere.

In summary, the twisted structure of a cloudy prominence was observed to undergo significant changes on 2021 February 10. A novel fluid instability mechanism involving the K\'arm\'an Vortex Street effects is proposed to explain the destabilization of the twisted prominence. This effect is manifested by the occurrence of successive vortex motion, accompanied by oscillations, resonance, and untwisting of the prominence. Consequently, the top structure of the prominence disintegrated, and the dark material embedded within it was transferred into the overlying magnetic arcades by merging. These findings contribute to the understanding of the structure and dynamic behavior of twisted prominences, and unveil a new mechanism for material transfer from prominences to the upper atmosphere.


\begin{acknowledgements}

The work is supported by the National Key R\&D Program of China (No. 2019YFA0405000, 2022YFF0503000, 2022YFF0503800, and 2021YFA1600500), the Strategic Priority Research Program of the Chinese Academy of Sciences (No. XDB41000000, XDB0560000), Key Research Program of Frontier Sciences of CAS (No. ZDBS-LY-SLH013), the National Natural Science Foundation of China (Nos. 12273061, 12350004, 12373111, 12063002, 12073038, and 12073042), and the Beijing Municipal Natural Science Foundation (No. 1202022).

 \end{acknowledgements}

%
%

\end{document}